\documentclass[12pt]{iopart}
\usepackage{fullpage,graphicx,epsf,ulem}
\usepackage{iopams}
\usepackage{amsfonts,amssymb}
\usepackage{amscd}
\usepackage{cite}
\usepackage{color}
\usepackage{graphicx}
\usepackage{epsfig}
\usepackage[english]{babel}
\usepackage{amsfonts}
\usepackage{soul}
\begin{document}
\def\be{\begin{equation}}
\def\ee{\end{equation}}
\def\bea{\begin{eqnarray}}
\def\eea{\end{eqnarray}}
\def\fr{\frac}
\def\l{\label}
\def\th{\theta}
\newenvironment{ale}{\color{blue}}

\newcommand{\dd}{\mbox{d}}
\newcommand{\gae}{\lower 2pt \hbox{$\, \buildrel {\scriptstyle >}\over {\scriptstyle
\sim}\,$}}
\newcommand{\lae}{\lower 2pt \hbox{$\, \buildrel {\scriptstyle <}\over {\scriptstyle
\sim}\,$}}
\title{Nonequilibrium inhomogeneous steady state distribution in disordered, mean-field rotator
systems}
\author{Alessandro Campa$^1$, Shamik Gupta$^{2}$, and Stefano Ruffo$^2$}
\address{$^1$ Complex Systems and Theoretical Physics Unit, Health and Technology
Department, Istituto Superiore di Sanit\`{a}, and INFN Roma1, Gruppo
Collegato Sanit\`{a}, Viale Regina Elena 299,
00161 Roma, Italy}
\address{$^2$ Dipartimento di Fisica e Astronomia and CSDC,
Universit\`{a} di Firenze, INFN and CNISM, via G. Sansone, 1 50019 Sesto
Fiorentino, Italy}
\ead{alessandro.campa@iss.infn.it,shamikg1@gmail.com,stefano.ruffo@gmail.com}

\begin{abstract}
We present a novel method to compute the phase space
distribution in the nonequilibrium stationary state of a wide class of mean-field
systems involving rotators subject to quenched disordered external drive
and dissipation. The method involves a series expansion of the stationary
distribution in inverse of the damping coefficient; the expansion coefficients satisfy recursion
relations whose solution requires computing a matrix where about three quarters of the elements 
vanish, making numerical evaluation simple and efficient. We illustrate our method for
the paradigmatic Kuramoto model of spontaneous collective
synchronization and for its two mode generalization, in presence of noise and inertia, and demonstrate an excellent
agreement between simulations and theory for the phase space distribution. 
\end{abstract}
\date{\today}
\maketitle
\tableofcontents
%---------------------------------------------------------------
\section{Introduction}
\l{secintro}
Spontaneous collective synchronization in a large population of coupled
oscillators of varying frequencies occurs in a wide variety of systems spanning  length and time scales of several orders
of magnitude. Examples are yeast cell suspensions \cite{Bier:2000}, cardiac cells \cite{Winfree:1980}, fireflies \cite{Buck:1988},
Josephson junctions \cite{Wiesenfeld:1998}, atomic recoil lasers
\cite{Javoloyes:2008}, animal flocks \cite{Ha:2010}, pedestrian motion on footbridges
\cite{Eckhardt:2007}, audience applause in concert halls
\cite{Neda:2000}, and many others \cite{Strogatz:2003}. The
paradigmatic minimal model to study synchrony and its emergence from asynchronous/incoherent
phase is the celebrated Kuramoto model, involving phase-only
oscillators of distributed natural frequencies coupled via a
mean field \cite{Kuramoto:1975,Kuramoto:1984}. The model enjoys the status of one of the most studied nonlinear dynamical
systems, which continues to provoke many unresolved
issues \cite{Strogatz-Acebron:2000-2005}. 

In the Kuramoto model, the oscillator phases follow a first-order
evolution in time. A generalized second-order stochastic dynamics, initially studied to
model better synchronization among flashing fireflies, accounts for the
finite moments of inertia of the oscillators (thus, the oscillators
become instead the rotators), and
for the stochastic fluctuations of the natural frequencies in time \cite{Ermentrout:1991,Acebron:1998,Acebron:2000}. The generalized
dynamics without stochasticity also arises in electrical power distribution networks \cite{Filatrella:2008,Rohden:2012}.

In a different context,
one may interpret the generalized Kuramoto 
dynamics as that of a system of interacting particles driven by quenched disordered external torques, and evolving
in presence of an external heat bath and dissipation; the Kuramoto model is recovered as the overdamped, noiseless
dynamics. This interpretation offers the very promising possibility
of studying the dynamics from 
statistical mechanical perspectives by using tools of kinetic
theory \cite{Gupta:2014,Gupta:2014a}. The
generalized model is a rich laboratory to study many-body
nonequilibrium dynamics in presence of external drive and quenched
disorder. The system relaxes at long times to a nonequilibrium stationary state (NESS)
\cite{Gupta:2014,Gupta:2014a}; the synchronized state (respectively, the incoherent state)
corresponds to a spatially inhomogeneous (respectively, spatially homogeneous) NESS.

Unlike in equilibrium, a NESS is characterized by a violation of
detailed balance, and an associated 
phase space distribution that cannot be expressed in the
Gibbs-Boltzmann form $\sim \exp(-\beta H)$ in terms of a Hamiltonian $H$ and an inverse temperature $\beta$. Instead,
the NESS distribution has to be obtained by studying on a
case-by-case basis the underlying dynamical model. This task
proves daunting especially for many-body interacting systems;
exact results are known only for simple models, often via {\it
tour de force} \cite{Derrida:1993}, while for more complex models, simulations and
approximation methods are invoked \cite{Privman:2005}.
 
In this paper, we 
provide a novel method to obtain the inhomogeneous NESS single-rotator phase
space distribution for the generalized Kuramoto
model in the thermodynamic limit. 
The resulting distribution has a nontrivial form with respect to the Gibbs-Boltzmann distribution in equilibrium. 
Our result provides an example 
of a computation of a nontrivial probability distribution
associated to a NESS in presence of quenched disorder. 
Significantly, our proposed method applies not just to the case
at hand, but, in fact, to {\it any} periodic two-body mean-field interaction
potential,
thereby providing a general framework to compute NESS distribution in a
wide class of systems. 

Our method involves a series expansion of the stationary
distribution in terms of the inverse of the
damping coefficient, as opposed to an expansion in the same parameter of
the Kramers operator for the time evolution of the single-rotator
distribution \cite{Risken:1996}. In our method, the expansion coefficients satisfy recursion
relations whose solution requires evaluation of a matrix where about three quarters of the elements 
vanish, making the numerical
implementation very efficient computationally and also simple compared
to the operator-expansion method. Also, in contrast to application of
the latter method to systems in external fields
\cite{Risken:1996}, we develop our method for mean-field systems that
require self-consistent evaluation of the mean fields. Note that the series expansion of the distribution
is asymptotic in nature (just as the expansion of the Kramers operator \cite{Risken:1996}), and one can
resort to the Borel summation method to sum the series properly \cite{Hardy:1991}.
In the context of the generalized Kuramoto dynamics, we demonstrate an excellent agreement of our theory
with simulation results for the phase space distribution. The latter is characterized by a
spatially non-uniform temperature profile distinctive of NESSs and
absent in equilibrium.

We remark that in a recent work \cite{Gupta:2014} we studied the generalized Kuramoto
dynamics, with a focus on obtaining the complete phase diagram of the
model for a general unimodal frequency distribution of the oscillators. 
Based on simulations and some analytic bounds, we demonstrated
that the model shows a nonequilibrium first-order phase transition from
a synchronized phase at low values of the relevant parameters to an
incoherent phase at high values. However, our analytical approach allowed to obtain only the homogeneous
NESS single-rotator phase space distribution and to study its dynamical stability, while we could not
compute analytically the distribution of the inhomogeneous NESS,
namely, the synchronized state. In the present work we do not address the issue of phase transitions,
but we focus on a method that computes the inhomogeneous NESS, not just for the Kuramoto potential,
but for any other periodic mean-field interaction potential. Thus, we also fill the gap left in our previous work.

The paper is organized as follows. In the following section, we state
the setting of the class of mean-field models that forms the object of
our study, and write the dynamical equations in a convenient dimensionless form. In Section
\ref{seckramers}, we discuss the Kramers equation for the time
evolution of the single-rotator phase space distribution, and in
particular, present in detail our method to compute its stationary
solution for the inhomogeneous phase. Some of the technical details are
relegated to the Appendix. In Section \ref{seckura}, we illustrate our method by
considering the representative case of the Kuramoto interaction
potential, and demonstrate an excellent agreement between theory and
simulations for several physically relevant observables. The paper ends
with conclusions.

\section{The class of models}
\l{secclass}
We now turn to deriving our results, by first stating the setting of the generalized Kuramoto dynamics: We have $N$ globally
coupled rotators of same moment of inertia $m$, with the $i$th rotator,
$i=1,2,\ldots,N$, having its natural frequency $\omega_i$ a quenched random variable given
by a common distribution ${\cal G}(\omega)$. The phases $\theta_i$'s
$\in [0,2\pi]$ and the angular velocities $v_i$'s follow the equations \cite{Acebron:1998,Acebron:2000}
\bea
&&\frac{\dd \theta_i}{\dd t}=v_i \nonumber \\
\l{inertialeq} \\ 
&&m \fr{\dd v_i}{\dd t} = -\gamma v_i + \gamma
\omega_i  -\frac{K}{N}\sum\limits_{j=1}^N \fr{\partial u(\th_j -
\th_i)}{\partial \th_i}+\sqrt{D}\eta_i(t), \nonumber
\eea
where $\gamma$ is
the damping coefficient, $K$ is the coupling constant that is scaled down by
$N$ to have a well-defined behavior of the associated term in the thermodynamic limit $N\to \infty$, $u(\theta_i-\theta_j)$ is
the two-body mean-field interaction potential \cite{note1}, while $\eta_i$ is a
Gaussian white noise:
\be
\langle \eta_i(t)\rangle = 0, \,\,\,\,\,\,\,\,\, \langle \eta_i(t)
\eta_j(t')\rangle = 2\delta_{ij}\delta(t-t').
\l{noisepar}
\ee
Here, the angular brackets denote averaging with respect to noise
realizations. The constant $D$ in Eq.~(\ref{inertialeq}) quantifies the strength of the noise force. Noting that $u(\th)$ is
periodic and even in $\th$ \cite{note2}, and taking $u(0)=0$ without loss of
generality, a Fourier expansion yields
\be
u(\th)=\sum_{s=1}^\infty \widetilde{u}_s[1-\cos(s\th)].
\l{potexpr}
\ee
The Kuramoto potential corresponds to the choice $\widetilde{u}_1=1,\widetilde{u}_{s >1}=0$.
The noiseless ($D=0$), overdamped ($m/\gamma \to 0$) limit
recovers the Kuramoto
dynamics \cite{Kuramoto:1975,Kuramoto:1984}:
\be
\fr{\dd \th_i}{\dd t}
= \omega_i+ \frac{\widetilde{K}}{N}\sum_{j=1}^N \sin(\th_j-\th_i),
\l{kuradyn}
\ee
where $\widetilde{K}\equiv K/\gamma$. 

Common to studies of the Kuramoto model, we consider a unimodal
${\cal G}(\omega)$, i.e., one which is symmetric about a single maximum
(same as the average $\langle \omega \rangle$), and with width $\sigma$.
The dynamics~(\ref{inertialeq}) is invariant under $\th_i \to \th_i
+ \langle \omega \rangle t, v_i \to v_i + \langle \omega \rangle,
\omega_i \to \omega_i + \langle \omega \rangle$, and the effects of
$\sigma$ may be made explicit by replacing $\omega_i$ in the second
equation with $\sigma \omega_i$. We consider from now on the
dynamics~(\ref{inertialeq}) with $\omega_i \to
\sigma \omega_i$, and take ${\cal G}(\omega)$ to have zero mean and unit width, without loss of
generality. 

On interpreting the model (\ref{inertialeq}) as that of interacting
particles, $m$ becomes the mass, $\th_i$ the angular coordinate for the
motion along a unit circle, $v_i$ the angular velocity, and $\gamma \omega_i$ the quenched disordered
external torque \cite{Gupta:2014,Gupta:2014a}.
Introduced to
mimic fluctuations of the natural frequencies \cite{Sakaguchi:1988}, the Gaussian
noise can also be interpreted as fluctuations due to coupling
to a heat bath at temperature $T$, and one may invoke the
fluctuation-dissipation relation to relate the strength of the noise to
the temperature $T$, as $D=\gamma k_B T$ \cite{note3}. In this case, Eq.~(\ref{inertialeq}) in the absence of
the $\omega_i$'s describes the dynamics of a Hamiltonian system in contact with a heat bath, and
the stationary state is in equilibrium, with probability of
configurations $\sim \exp(-H/T)$, where $H$ is the Hamiltonian
\be
H=\sum_{i=1}^N \frac{p_i^2}{2m}+\frac{K}{2N}\sum_{i,j=1}^Nu(\th_i-\th_j),
\l{hamilfun}
\ee
with $p_i\equiv mv_i$ the angular momentum of the $i$-th particle. In presence of
$\omega_i$'s, the dynamics relaxes to a NESS \cite{Gupta:2014,Gupta:2014a}.

It is convenient for further analysis to write Eq.~(\ref{inertialeq}) in
a dimensionless form. Introducing dimensionless variables 
\bea
&&\overline{t}\equiv  t\sqrt{K/m}, \\
&&\overline{v}_i\equiv v_i\sqrt{m/K}, \\
&&1/\sqrt{\overline{m}}\equiv \gamma/\sqrt{Km}, \\
&&\overline{\sigma} \equiv \gamma\sigma/K, \\
&&\overline{T} \equiv T/K,\\
&&\overline{\eta}_i(\overline{t})\equiv \eta_i(t)\sqrt{\gamma T}/K,
\eea
Eq.~(\ref{inertialeq}) becomes
\bea
\frac{\dd \th_i}{\dd \overline t}&=&\overline{v}_i, \nonumber \\
\l{eom-scaled} \\ 
\frac{\dd \overline{v}_i}{\dd \overline{t}}
&=&\frac{-\overline{v}_i}{\sqrt{\overline{m}}}+\sum\limits_{s=1}^\infty
s \widetilde{u}_s R_s
\sin(s\psi-s\th_i)+\overline{\sigma}\omega_i+\Big(\fr{\overline{T}}{\sqrt{\overline{m}}}\Big)^{1/2}\overline{\eta}_i(\overline{t}),
\nonumber
\eea
where
\be
R_s(t) e^{is\psi(t)} \equiv \frac{1}{N} \sum_{j=1}^N e^{is \th_j(t)},
\l{deforderpar}
\ee
and the noise satisfies 
\be
\langle \overline{\eta}_i(\overline{t})\overline{\eta}_j(\overline{t}')\rangle =
2\delta_{ij}\delta(\overline{t}-\overline{t}').
\ee

For $m = 0$, using dimensionless time 
\be
\overline{t}\equiv t(K/\gamma),
\ee
the dynamics (\ref{inertialeq}) becomes the overdamped motion
\be
\frac{\dd \th_i}{\dd \overline{t}} = \sum_{s=1}^\infty s \widetilde{u}_s R_s
\sin(s\psi-s\th_i)+\overline{\sigma}\omega_i+\sqrt{\overline{T}}\overline{\eta}_i(\overline{t}),
\l{overdampeq}
\ee
which in the case $\widetilde{u}_1=1, \widetilde{u}_{s>1}=0$ and at
$\overline{T}=0$ becomes the Kuramoto
dynamics. Associated with the $s$th Fourier mode of the interaction potential is
the magnitude of the mean field $R_s$ acting on one rotator due to its interaction with all the others, while
$s\psi$ is the corresponding phase. In particular, $R_1$ measures complete phase coherence among all the rotators, and its stationary
value $R_1^{\rm st} \equiv R_1(t \to \infty)$ serves as the synchronization order parameter. From now on, we consider the dynamics
(\ref{eom-scaled}) that, besides $N$, depends on the parameters $\overline{m}$,
$\overline{T}$, and $\overline{\sigma}$; we will drop the overbars for
notational simplicity. Note that dynamics (\ref{eom-scaled}) reduces to
dynamics (\ref{overdampeq}) in the overdamped limit.

In the next section, we discuss the Kramers equation for the time
evolution of the single-rotator phase space distribution, and in
particular, our method to compute its stationary solutions.

\section{The Kramers equation and its stationary solutions}
\l{seckramers}
In the thermodynamic limit, the dynamics~(\ref{eom-scaled}) is described by the single-rotator phase space distribution
$f(\th,v,\omega,t)$, giving at time $t$ for each $\omega$ the distribution probability
of rotators with phase $\th$ and angular velocity $v$. We have 
\be
f(\th,v,\omega,t)=f(\th+2\pi,v,\omega,t),
\ee
and the normalization 
\be
\int \dd \th \dd v~ f(\th,v,\omega,t)=1 ~~\forall~\omega.
\ee

The time evolution of $f$, obtained by truncating to
lowest order in $1/N$ the Bogoliubov-Born-Green-Kirkwood-Yvon (BBGKY) hierarchy
equations, follows the Kramers equation
\cite{Gupta:2014}
\be
\fl
\l{Kramerst}
\fr{\partial f}{\partial t}=-v\fr{\partial f}{\partial \th}
+\fr{T}{\sqrt{m}}\fr{\partial^2 f}{\partial v^2} +\fr{\partial}{\partial v}
\left[ \left( \frac{v}{\sqrt{m}}-\sum_{s=1}^\infty s \widetilde{u}_s R_s
\sin(s\psi-s\th)-\sigma\omega \right) f \right],
\ee
where
\be
R_s(t)e^{is\psi(t)}=\int \dd \th \dd v \dd \omega ~{\cal G}(\omega)e^{is\th}f(\th,v,\omega,t).
\l{deforderkram}
\ee
We are interested in the stationary state solutions of the Kramers equation, obtained by setting the
left hand side of Eq.~(\ref{Kramerst}) to $0$. As already mentioned, the
stationary state is a NESS, unless $\sigma=0$. In the stationary state $R_s$ and $\psi$ are time independent;
$\psi$ can be set equal to $0$ by redefining the origin of the $\th_i$'s, while with $R_s^{\rm st}$ we will
denote the stationary value of $R_s$. We will also use the definition
\be
G(\th) \equiv \sum_{s=1}^\infty s \widetilde{u}_s R_s^{\rm st} \sin(s\th).
\l{defgtheta}
\ee
The $\theta$-independent solution characterizing the incoherent phase, for which
$R_s^{\rm st} = 0 ~\forall~ s$ and thus $G(\theta)=0$, is given by \cite{Acebron:2000}:
\be
f^{\rm inc}(\th,v,\omega)=\frac{1}{2\pi}\sqrt{\frac{1}{2\pi T}}
\exp \left[-\frac{(v-\sigma \omega \sqrt{m})^2}{2T}\right] .
\l{incsol}
\ee
The synchronized phase distribution for $\sigma=0$ is given by the Gibbs-Boltzmann measure
$\sim \exp [-v^2/(2T)-\int \dd \theta G(\theta)]$; for general $\sigma$,
we expand it as 
\be
f^{\rm syn}(\th,v,\omega)=
\Phi_0\left( \fr{v}{\sqrt{2T}}\right) \sum_{n=0}^\infty b_n(\theta,\omega)
\Phi_n\left( \fr{v}{\sqrt{2T}}\right),
\l{f-expansion}
\ee
where the functions $b_n$'s satisfy
$b_n(\th,\omega)=b_n(\th+2\pi,\omega)$, while $\Phi_n(ax)$ is the Hermite function:
\be
\Phi_n(ax)=\sqrt{\frac{a}{2^n n!\sqrt{\pi}}}\exp\left[-\frac{a^2x^2}{2}\right]H_n(ax),
\l{hermfunct}
\ee
with $H_n(x)$'s being the $n$-th degree Hermite polynomial. The functions
$\Phi_n$'s are orthonormal: $\int \dd x ~\Phi_m(ax) \Phi_n(ax) = \delta_{mn}$.
Normalization of $f^{\rm syn}(\th,v,\omega)$ implies that $\int_0^{2\pi} \dd \th ~b_0(\th,\omega)=1$,
while the self-consistent values of the parameters $R_s^{\rm st}$ are given by
\be
R_s^{\rm st}=\int \dd \omega~{\cal G}(\omega)\int_0^{2\pi}\dd \theta~b_0(\th,\omega) e^{is\th}.
\l{rstself}
\ee
Furthermore, using $\int \dd x~x\Phi_0(ax)\Phi_n(ax)=1/(\sqrt{2}a)\delta_{n,1}$, 
we obtain that
\be
\int \dd v~vf^{\rm syn}(\th,v,\omega) = \sqrt{T}b_1(\th,\omega).
\l{eqcurrent}
\ee
On the other hand, integrating over $v$ the stationary state Kramers equation, we obtain that
$\int \dd v~vf^{\rm syn}(\th,v,\omega)$ and, hence, $b_1(\th,\omega)$, does not depend on $\theta$.

The choice of the Hermite functions in the expansion
(\ref{f-expansion}) is motivated by the fact for $\sigma=0$, the
distribution $f_{\rm syn}(\th,v,\omega)$ should have the Gibbs-Boltzmann
form, $f_{\rm syn}(\th,v,\omega) \sim \exp [-v^2/(2T)-\int {\rm d} \theta G(\theta)]$. As will be
shown later, the expansion coefficients $b_n$ for this case satisfy
$b_0(\th,0) \sim \exp [-\int {\rm d} \theta G(\theta)], ~b_n(\th,0)=0$ for $n>0$, so that only
the $n=0$ term in the expansion (\ref{f-expansion}) needs to be taken into
account; then, with $\Phi_0(x)\sim\exp(-x^2/2)$, the product
$\Phi_0\left( \fr{v}{\sqrt{2T}}\right)\Phi_0\left( \fr{v}{\sqrt{2T}}\right)$
appearing in the expansion correctly
reproduces the velocity-part of the distribution $\sim \exp [-v^2/(2T)]$.

Plugging the expansion (\ref{f-expansion}) into the stationary state Kramers
equation, using the known recursion relations for the Hermite
polynomials, and equating to zero the coefficient of each $\Phi_n$, we
get
\bea
\fl
\sqrt{nT}\fr{\partial b_{n-1}(\th,\omega)}{\partial \theta}
+\sqrt{(n+1)T}\fr{\partial b_{n+1}(\th,\omega)}{\partial \theta}+\fr{n}{\sqrt{m}}b_n(\th,\omega)
+\sqrt{\fr{n}{T}}b_{n-1}(\th,\omega)[G(\th)-\sigma \omega]=0 \nonumber \\
\l{bp-eqn}
\eea
for $n=0,1,2,\dots$ (with the understanding that $b_{-1}(\th,\omega)\equiv 0$). Equation~(\ref{bp-eqn}) is
a time-independent generalized version of the Brinkman's
hierarchy \cite{Brinkman:1956,Risken:1996,Durang:2013}.
The hierarchy was introduced to study the approach to a stationary state of a system of noninteracting 
particles subjected to external forces and noise. We remark that in our case we have a system of interacting
particles, so that the forces are both external (the driving torques) and due to an interaction potential.  
The equation for $n=0$ recovers the result that
$b_1(\th,\omega)$ is independent of $\theta$. Noting the scaling of
the various terms in Eq.~(\ref{bp-eqn}) with $m$, we expand $b_n(\th,\omega)$
as
\be
b_n(\th,\omega)=\sum_{k=0}^\infty (\sqrt{m})^k c_{n,k}(\th,\omega),
\l{bexpansion}
\ee
with $b_1(\th,\omega)$ independent of $\th$ implying that so is $c_{1,k}(\th,\omega)~\forall~k$.
The only constraint on $b_0(\th,\omega)$ being $\int_0^{2\pi} \dd \th ~b_0(\th,\omega)=1$, we may
without loss of generality choose $c_{0,k \ge 1}(0,\omega)=0$. This will
prove very useful for further analysis, as will be shown below. We now use Eq.~(\ref{bexpansion}) in Eq.~(\ref{bp-eqn}) and equate to zero the coefficient
of each power of $\sqrt{m}$. The term proportional to $\left(\sqrt{m}\right)^{-1}$ gives simply
\be\label{sup_m1}
nc_{n,0}(\th,\omega)=0, 
\ee
which implies that $c_{n,0}(\th,\omega)=0$ for $n>0$. The coefficient of the
term proportional to $\left(\sqrt{m}\right)^k$ leads to
\bea
\fl
\sqrt{nT}\fr{\partial c_{n-1,k}(\th,\omega)}{\partial \theta}
+\sqrt{(n+1)T}\fr{\partial c_{n+1,k}(\th,\omega)}{\partial \theta}
+\sqrt{nT} a(\th,\omega) c_{n-1,k}(\th,\omega) + nc_{n,k+1}(\th,\omega)=0 \nonumber \\
\l{finalsystem}
\eea
for $n,k=0,1,2,\ldots$ (with $c_{-1,k}(\th,\omega)\equiv 0$),
where $a(\th,\omega)\equiv[G(\th)-\sigma \omega]/T$. The system of equations (\ref{finalsystem})
can be solved recursively, as we detail now.

\subsection{Solution of the system of equations~(\ref{finalsystem})}
Let us first consider Eq.~(\ref{finalsystem}) for $n=0$; we immediately obtain that $c_{1,k}(\th,\omega)$
is independent of $\th$ for each $k$, as we had already inferred above. To proceed,
we consider Eq.~(\ref{finalsystem}) for $k=0$ and $n=2,3,\dots$; since $c_{n,0}(\th,\omega)=0$ for $n>0$,
we get that $c_{n,1}(\th,\omega)=0$ for $n>1$. Considering next the equation for $k=1$ and
$n=3,4,\dots$, we get $c_{n,2}(\th,\omega)=0$ for $n>2$; for $k=2$ and $n=4,5,\dots$
we get $c_{n,3}(\th,\omega)=0$ for $n>3$, and so on. We thus arrive at the general
result that $c_{n,k}(\th,\omega)=0~\forall~k<n$. In Fig. \ref{flowfig}, we display the coefficients
$c_{n,k}$ in a matrix. The result just obtained shows that the matrix is upper triangular.
We are thus left to consider Eq.~(\ref{finalsystem}) for $n=1,2,\dots$ and $k \ge n-1$, or, equivalently,
for $k=0,1,2,\dots$ and $n=1,2,\dots,k+1$. In what follows, we will obtain the elements of the main
diagonal, $c_{n,n}(\th,\omega)$, then the elements of the first upper diagonal, $c_{n,n+1}(\th,\omega)$,
the elements of the second upper diagonal, $c_{n,n+2}(\th,\omega)$, and so on.
\begin{figure}
\centering
\includegraphics[scale=0.3]{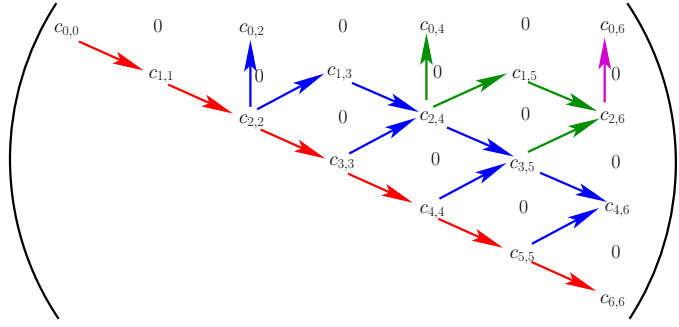}
\caption{Flow diagram for the evaluation of the expansion coefficients
$c_{n,k}(\th,\omega); n,k=0,1,2,\ldots,6$. Starting from the main diagonal,
arrows and different colors denote subsequent flows (see text). The elements below the main diagonal are all zero.}
\l{flowfig}
\end{figure}
Thus, let us begin by studying the case $n=1$ and $k=0$. We have
\be
\sqrt{T}\fr{\partial c_{0,0}(\th,\omega)}{\partial \theta}
+\sqrt{2T}\fr{\partial c_{2,0}(\th,\omega)}{\partial \theta}
+\sqrt{T} a(\th,\omega) c_{0,0}(\th,\omega) + c_{1,1}(\omega)=0.
\label{finalsystem_n1k0}
\ee
In this equation, $c_{2,0}(\th,\omega)=0$, while $c_{1,1}(\omega)$ is independent of $\th$.
We thus have a first-order differential equation for $c_{0,0}(\th,\omega)$, with
an unknown constant. The requirement of periodicity, i.e., $c_{0,0}(\th,\omega)=c_{0,0}(\th +2\pi,\omega)$,
fixes the value of this constant, and we obtain
\bea
c_{0,0}(\th,\omega)&=&c_{0,0}(0,\omega)e^{- g(\th,\omega)}\left[1+\left(e^{g(2\pi,\omega)}-1\right)
\fr{\int_0^\th \dd \th' e^{g(\th',\omega)}}{\int_{0}^{2\pi} \dd \th' e^{g(\th',\omega)}}\right],
\label{solc00} \\
c_{1,1}(\omega)&=&\sqrt{T}\fr{c_{0,0}(0,\omega)\left(1-e^{g(2\pi,\omega)}\right)}
{\int_{0}^{2\pi} \dd \th' e^{g(\th',\omega)}},
\label{solc11}
\eea
where $g(\th,\omega)=\int_0^\th \dd \th'a(\th',\omega)$, and $c_{0,0}(0,\omega)$ is to be fixed at
the end by the normalization of $b_0(\th,\omega)$. Having determined
$c_{0,0}(\th,\omega)$ and $c_{1,1}(\omega)$, we then obtain recursively the main diagonal
elements, by considering Eq.~(\ref{finalsystem}) for $n=2,3,\dots$ and $k=n-1$; this gives
\bea
\sqrt{nT}\fr{\partial c_{n-1,n-1}(\th,\omega)}{\partial \theta}
+\sqrt{(n+1)T}\fr{\partial c_{n+1,n-1}(\th,\omega)}{\partial \theta} \nonumber \\
+\sqrt{nT} a(\th,\omega) c_{n-1,n-1}(\th,\omega) + nc_{n,n}(\th,\omega)=0.
\label{finalsystem_nn}
\eea
Since $c_{n+1,n-1}(\th,\omega)=0$, we have
\be\label{solcnn}
c_{n,n}(\th,\omega)=-\sqrt{\fr{T}{n}}\left[\fr{\partial c_{n-1,n-1}(\th,\omega)}{\partial \theta}
+ a(\th,\omega )c_{n-1,n-1}(\th,\omega) \right]
\ee
for $n=2,3,\dots$. In particular, for $n=2$ the first term within the
square brackets is absent, since $c_{1,1}(\omega)$
is independent of $\th$. We note that all the functions $c_{n,n}(\th,\omega)$ are
proportional to $c_{0,0}(0,\omega)$. 

We now determine the elements of the first upper diagonal. We consider
Eq.~(\ref{finalsystem}) for $n=1$ and $k=1$: 
\be
\sqrt{T}\fr{\partial c_{0,1}(\th,\omega)}{\partial \theta}
+\sqrt{2T}\fr{\partial c_{2,1}(\th,\omega)}{\partial \theta}+\sqrt{T} a(\th,\omega)
c_{0,1}(\th,\omega) + c_{1,2}(\omega)=0.
\label{finalsystem_n1k1}
\ee
This equation has exactly the same structure as Eq.~(\ref{finalsystem_n1k0}), since
$c_{2,1}(\th,\omega)=0$, and $c_{1,2}(\omega)$ is a constant independent of $\th$. At this point,
we use the fact that $c_{0,k}(0,\omega)=0$ for $k\ge 1$. Then, the solution of Eq.~(\ref{finalsystem_n1k1})
is simply $c_{0,1}(\th,\omega)=c_{1,2}(\omega)\equiv 0$. Next,
by considering Eq.~(\ref{finalsystem}) for
$n=2,3,\dots$ and $k=n$, and proceeding similarly, we obtain that all the functions
$c_{n,n+1}(\th,\omega)$, i.e., the elements of the first upper diagonal of Fig. \ref{flowfig}, vanish.

Next, we determine the elements of the second upper diagonal, beginning by considering
Eq.~(\ref{finalsystem}) for $n=1$ and $k=2$:
\be
\sqrt{T}\fr{\partial c_{0,2}(\th,\omega)}{\partial \theta}
+\sqrt{2T}\fr{\partial c_{2,2}(\th,\omega)}{\partial \theta}+\sqrt{T} a(\th,\omega) 
c_{0,2}(\th,\omega) + c_{1,3}(\omega)=0. 
\label{finalsystem_n1k2}
\ee
In this equation, $c_{2,2}(\th,\omega)$ is known from Eq.~(\ref{solcnn}). Then, from the requirement of periodicity of
$c_{0,2}(\th,\omega)$, and using $c_{0,2}(0,\omega)=0$, we obtain the solutions
\bea\label{solc02}
c_{0,2}(\th,\omega)&=&\sqrt{2}\fr{\int_0^{2\pi} \dd \th' \fr{\partial c_{2,2}(\th',\omega)}{\partial \theta'}
e^{g(\th',\omega)}}{\int_0^{2\pi} \dd \th' e^{g(\th',\omega)}}e^{-g(\th,\omega)}\int_0^\th \dd \th'
e^{g(\th',\omega)}\nonumber \\ &&-\sqrt{2}e^{-g(\th,\omega)}\int_0^\th \dd \th' \fr{\partial c_{2,2}(\th',\omega)}
{\partial \theta'} e^{g(\th',\omega)}, \\
c_{1,3}(\omega)&=&-\sqrt{2T}\fr{\int_0^{2\pi} \dd \th' \fr{\partial c_{2,2}(\th',\omega)}{\partial \theta'}
e^{g(\th',\omega)}}{\int_0^{2\pi} \dd \th' e^{g(\th',\omega)}}.
\label{solc13}
\eea
Again, these functions are proportional to $c_{0,0}(0,\omega)$. Having
determined $c_{0,2}$ and $c_{1,3}$, we obtain recursively the elements
of the second upper diagonal, i.e., the functions $c_{n,n+2}$,
from Eq.~(\ref{finalsystem}) by considering $n=2,3,\dots$ and $k=n+1$:
\bea
\sqrt{nT}\fr{\partial c_{n-1,n+1}(\th,\omega)}{\partial \theta}
+\sqrt{(n+1)T}\fr{\partial c_{n+1,n+1}(\th,\omega)}{\partial \theta}\nonumber \\
+\sqrt{nT} a(\th,\omega) c_{n-1,n+1}(\th,\omega)
+ nc_{n,n+2}(\th,\omega)=0. 
\label{finalsystem_nnp2}
\eea
With the main diagonal elements already determined, this gives
\bea
c_{n,n+2}(\th,\omega)&=&-\sqrt{\fr{T}{n}}\left[ \fr{\partial c_{n-1,n+1}(\th,\omega)}{\partial \theta}+
a(\th,\omega)c_{n-1,n+1}(\th,\omega)\right] \nonumber \\
&&-\fr{\sqrt{(n+1)T}}{n}\fr{\partial c_{n+1,n+1}(\th,\omega)}{\partial \theta}, 
\label{solcnnp2}
\eea
for $n=2,3,\dots$. In particular, for $n=2$, the first term within
the square brackets is absent as $c_{1,3}(\omega)$ is independent of $\th$. Also these
functions are proportional to $c_{0,0}(0,\omega)$.

Next, we show that the elements of the third upper diagonal vanish.
Considering now Eq.~(\ref{finalsystem}) for $n=1$ and $k=3$, we have
\be
\sqrt{T}\fr{\partial c_{0,3}(\th,\omega)}{\partial \theta}
+\sqrt{2T}\fr{\partial c_{2,3}(\th,\omega)}{\partial \theta}+\sqrt{T}
a(\th,\omega) c_{0,3}(\th,\omega) + c_{1,4}(\omega)=0. 
\label{finalsystem_n1k3}
\ee
In this equation, $c_{2,3}$ has been previously determined to be
vanishing identically, so that the solution of the
last equation is simply $c_{0,3}(\th,\omega)=c_{1,4}(\omega)\equiv 0$. Then,
considering Eq.~(\ref{finalsystem}) for
$n=2,3,\dots$ and $k=n+2$, we find that all the elements of the third upper diagonal, $c_{n,n+3}$,
vanish. 

At this point, the procedure of determining the coefficients $c_{n,k}$'s should be clear. All the elements
of the upper diagonals of odd order vanish, this being equivalent to the fact that in the portion of each
row above the main diagonal, one element every two vanishes, i.e.,
$c_{n,n+1+2k}\equiv 0$ for $n,k=0,1,2,\dots$. All the nonvanishing
elements are proportional to $c_{0,0}(0,\omega)$. The expressions for the main
diagonal elements have already been given in Eqs.~(\ref{solc00}), (\ref{solc11}) and
(\ref{solcnn}). On the basis of the analysis above, we can write down the general expressions for
the nonvanishing non-diagonal elements as 
\bea
c_{0,2k}(\th,\omega)&=&\sqrt{2}\fr{\int_0^{2\pi} \dd \th' \fr{\partial c_{2,2k}(\th',\omega)}{\partial \theta'}
e^{g(\th',\omega)}}{\int_0^{2\pi} \dd \th' e^{g(\th',\omega)}}e^{-g(\th,\omega)}\int_0^\th \dd \th' e^{g(\th',\omega)} \nonumber \\
&&-\sqrt{2}e^{-g(\th,\omega)}\int_0^\th \dd \th' \fr{\partial c_{2,2k}(\th',\omega)}{\partial \theta'}
e^{g(\th',\omega)}, 
\l{solcn02k} \\
c_{1,1+2k}(\omega)&=&-\sqrt{2T}\fr{\int_0^{2\pi} \dd \th' \fr{\partial c_{2,2k}(\th',\omega)}{\partial \theta'}
e^{g(\th',\omega)}}{\int_0^{2\pi} \dd \th' e^{g(\th',\omega)}},
\l{solcn11p2k} \\
c_{2,2+2k}(\th,\omega)&=& -\sqrt{\fr{T}{2}} a(\th,\omega)
c_{1,1+2k}(\omega)-\fr{\sqrt{3T}}{2}\fr{\partial c_{3,1+2k}(\th,\omega)}{\partial \theta},
\l{solcn22p2k} \\ 
c_{n,n+2k}(\th,\omega)&=&-\sqrt{\fr{T}{n}}\left[ \fr{\partial c_{n-1,n-1+2k}(\th)}{\partial \theta}
+a(\th,\omega)c_{n-1,n-1+2k}(\th,\omega)\right] \nonumber \\
&&-\fr{\sqrt{(n+1)T}}{n}\fr{\partial c_{n+1,n-1+2k}(\th,\omega)}{\partial
\theta} \,\,\,\,\,\, n \ge 3, 
\l{solcnnp2k}
\eea
with $k=1,2,\dots$.

We show schematically in Fig.~\ref{flowfig} the flow of the solution up
to $n=k=6$, while that for higher values proceeds analogously. As shown,
the system (\ref{finalsystem}) computes progressively each element of the main
diagonal, and then the elements of the second upper diagonal,
each one determined by the knowledge of two previously determined
elements, and so on. Each element of the matrix is proportional to
$c_{0,0}(0,\omega)$, which is fixed by the normalization of
$f^{\rm syn}$: $\sum_{k=0}^\infty \int_0^{2\pi} \dd \th~(\sqrt{m})^{2k} c_{0,2k}(\th,\omega)=1$.
The values of $R_s^{\rm st}$ have to be determined self-consistently.

We end the section by pointing out that for $\sigma = 0$, or equivalently for $\omega = 0$,
the equilibrium Gibbs-Boltzmann distribution is simply recovered in our procedure. In fact,
it is immediate to see that for $\omega = 0$ the solution of (\ref{bp-eqn}) is
$b_0(\th,0) \sim \exp [-\int \dd \theta G(\theta)]$ and $b_n(\th,0)=0$ for $n>0$.
Coherently with this, the solution of (\ref{finalsystem}) for $\omega = 0$ is
$c_{0,0}(\th,0) \sim \exp [-\int \dd \theta G(\theta)]$, with all the other
$c_{n,k}(\th,0)=0$ vanishing. This can be readily obtained from
Eqs. (\ref{solc00})-(\ref{solcnnp2k}) by the fact that $g(2\pi,0)=0$.

\section{Illustration for a representative example: The Kuramoto
interaction potential}
\l{seckura}
In order to illustrate an implementation of our method, we now apply it to the Kuramoto potential.
In this case, as noted above, only the first Fourier term with $s=1$ needs to be taken
into account in the interaction potential. For illustrative purpose, let us choose
a representative ${\cal G}(\omega)$, namely, a Gaussian:
${\cal G}(\omega)=1/(\sqrt{2\pi})\exp(-\omega^2/2)$, and study two physically relevant quantities
in the synchronized phase. One is the marginal $\th$-distribution,
$n(\th) \equiv \int_{-\infty}^\infty
\dd \omega~{\cal G}(\omega)\int_{-\infty}^\infty \dd v~f^{\rm syn}(\th,v,\omega)$,
i.e., the density profile; using the orthonormality of the Hermite functions, one gets
\be
n(\th)=\int_{-\infty}^\infty d\omega~{\cal G}(\omega)b_0(\th,\omega).
\l{exprdens}
\ee
The other is the quantity
$p(\theta) \equiv \int_{-\infty}^\infty
\dd \omega~{\cal G}(\omega)\int_{-\infty}^\infty \dd v~v^2f^{\rm syn}(\th,v,\omega)$,
which is proportional to the local pressure \cite{Huang}. Using again the
orthonormality of the Hermite functions, one has
\be 
p(\th) = T\int_{-\infty}^\infty d\omega~{\cal G}(\omega)\left(\sqrt{2}b_2(\th,\omega)+b_0(\th,\omega)\right).
\l{exprpres}
\ee
We thus need the coefficients $b_0(\th,\omega)$ and $b_2(\th,\omega)$, whose
evaluation requires truncating the expansion (\ref{bexpansion}) at
suitable values $k_{\rm trunc}$ of $k$. From Fig. \ref{flowfig}, we see that knowing
$c_{2,2k}$ allows to compute $c_{0,2k}$, so it is natural to choose the same
$k_{\rm trunc}$ for both $b_0(\th,\omega)$ and $b_2(\th,\omega)$.

In Fig.~\ref{fig-ness-th}, we demonstrate an excellent agreement between theory and simulations for the density
$n(\th)$, for given values of $(m,T,\sigma)$. The simulations are performed through
integration of the $2N$ equations of motion (\ref{eom-scaled}) by using the algorithm of
Ref. \cite{Gupta:2014} and timestep $\delta t=0.01$. From the plots of the figure, it is evident that our
analytical approach works very well for both small and large values of $m$.
\begin{figure}
\centering
\includegraphics[width=75mm]{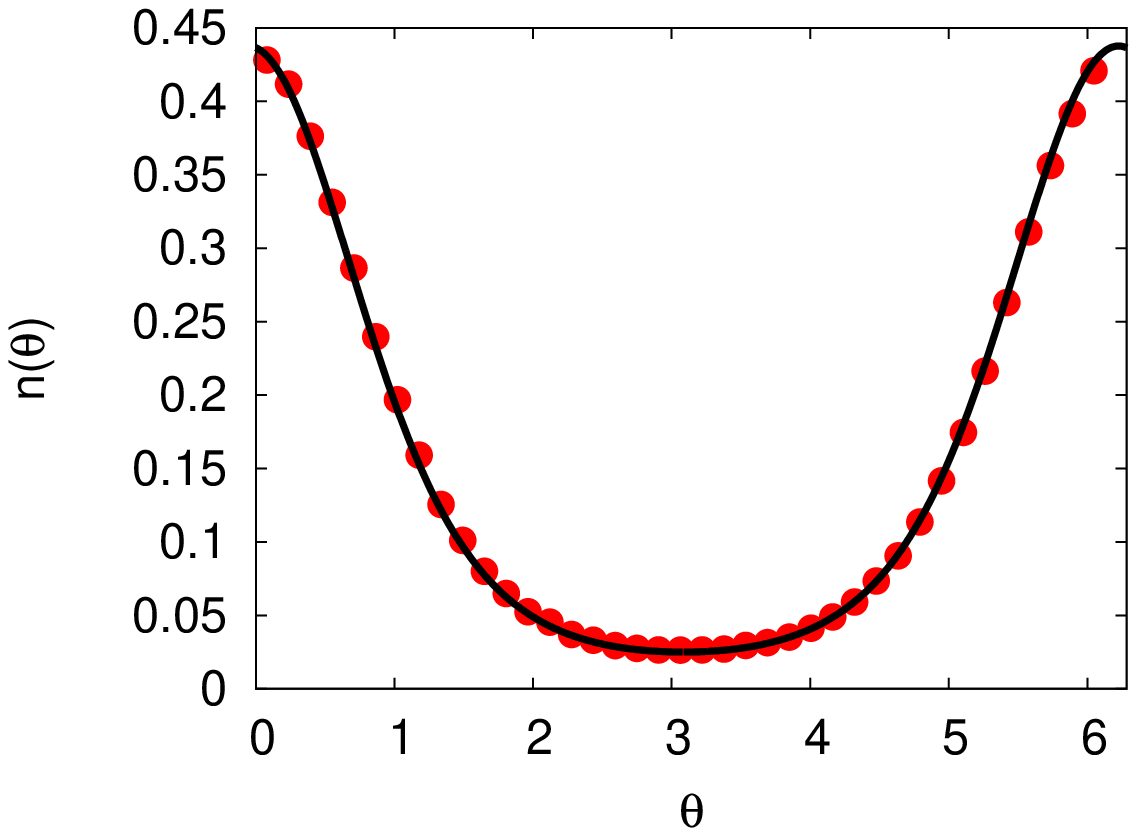}
\includegraphics[width=75mm]{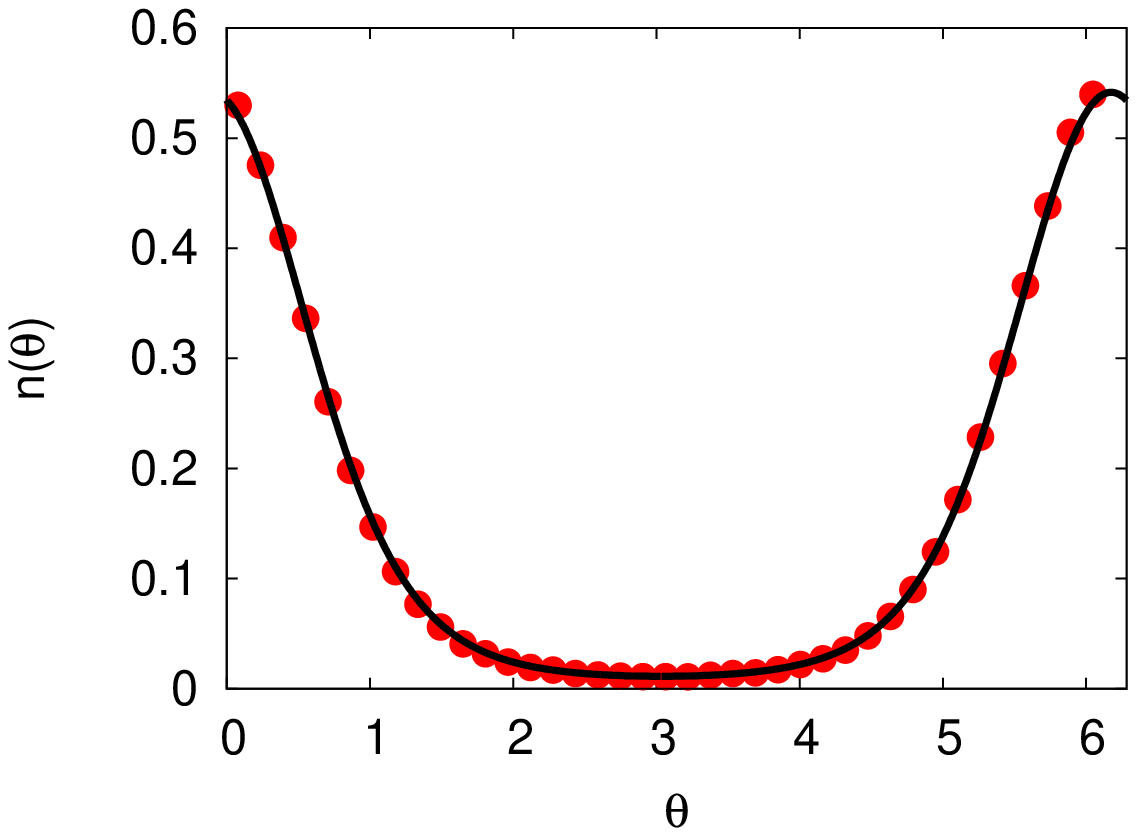}
\caption{Density $n(\theta)$ in the dynamics~(\ref{eom-scaled}) with
$\widetilde{u}_1=1$, $\widetilde{u}_{s>1}=0$, a Gaussian ${\cal G}(\omega)$,
for $m=0.25$, $T=0.25$, $\sigma=0.295$, $k_{\rm trunc}=12$ (left panel), and
for $m=5.0$, $T=0.25$, $\sigma=0.2$, $k_{\rm trunc}=2$ (right panel).
Simulations (points) are for $N=10^6$; the theoretical predictions
are denoted by lines.}
\l{fig-ness-th}
\end{figure}
The agreement is confirmed also in the plot of the quantity $p(\th)$, proportional to the local pressure,
as shown in the left panel of Fig.~\ref{fig-ness-temp}, where the parameters are the same of those in the
left panel of Fig.~\ref{fig-ness-th}.
\begin{figure}
\centering
\includegraphics[width=75mm]{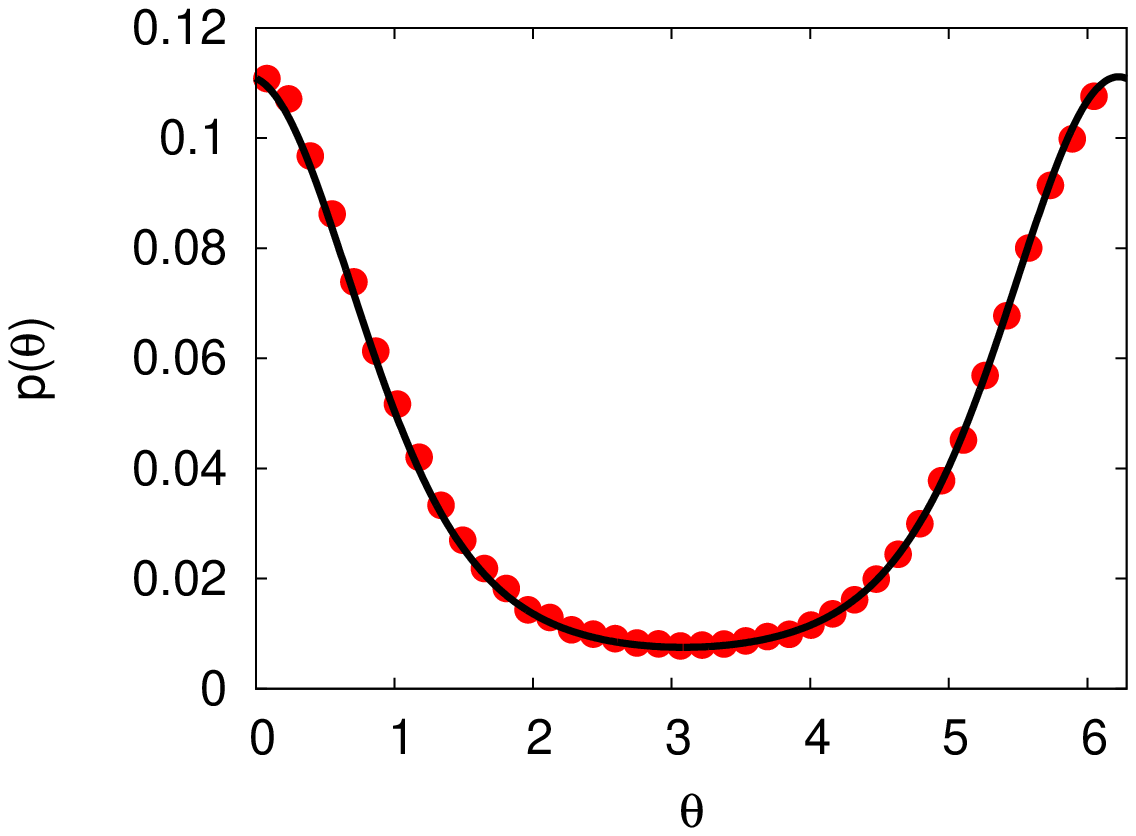}
\includegraphics[width=75mm]{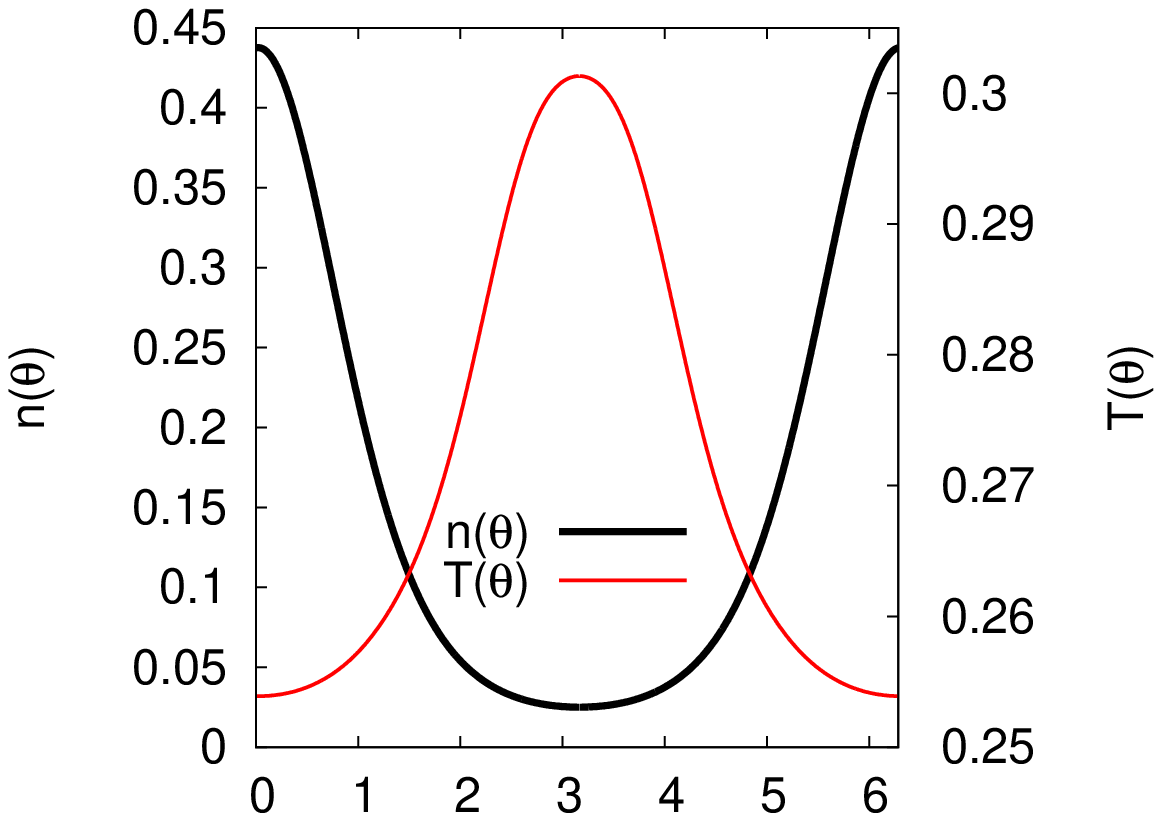}
\caption{The left panel shows the pressure $p(\th)$ for the same parameters as in the left panel of
Fig.~\ref{fig-ness-th}. Simulations (points) are for $N=10^6$; the theoretical predictions
are denoted by lines. The right panel shows the local temperature $T(\th)=p(\th)/n(\th)$ and its
anticorrelation with the density $n(\th)$.}
\l{fig-ness-temp}
\end{figure}
As a further demonstration of the validity of our method, we list in
Table~\ref{table1} the value of $R_1^{\rm st}$ obtained in theory and
simulations ($N=10^6$) for several $\sigma$'s and $m=0.25,T=0.25$; again, we observe a very good
agreement, within numerical accuracies.

\begin{table}
\centering
\begin{tabular}{|c|c|c|c|c|c|c|}  \hline
$\sigma$                    &$0.0$   & $0.05$  & $0.1$   & $0.15$  & $0.2$ & $0.25$ \\ \hline 
$R_1^{\rm st}$ (Theory)      &$0.829$ & $0.825$ & $0.813$ & $0.789$ & $0.75$ & $0.686$  \\ \hline
$R_1^{\rm st}$ (Simulations) &$0.829$ & $0.825$ & $0.812$ & $0.787$ &
$0.747$ & $0.680$ \\ \hline 
\end{tabular}
\caption{$R_1^{\rm st}$ vs. $\sigma$ obtained in theory and
simulations in the dynamics~(\ref{eom-scaled}), with
$\widetilde{u}_1=1,\widetilde{u}_{s>1}=0$, a Gaussian ${\cal
G}(\omega)$, for several values of $\sigma$ at $m=0.25,T=0.25,k_{\rm
trunc}=12$.}
\l{table1}
\end{table}

The ratio $p(\th)/n(\th)$ gives the temperature $T(\th)$. In equilibrium, one has a spatially uniform
temperature profile, i.e., $T(\th)$ equals the temperature $T$ of the
heat bath, independent of $\th$. Then, the spatially non-uniform
temperature profile in the right panel of Fig.~\ref{fig-ness-temp} (where we show the theoretical computation),
is a further demonstration that the synchronized state is a NESS.
The panel also depicts a density-temperature anticorrelation, i.e.,
the temperature peaks at a $\th$ at which the density
is minimum, and vice versa. This phenomenon of temperature inversion
occurs in inhomogeneous plasmas (e.g., the
Solar corona \cite{Golub:2009}, interstellar molecular clouds
\cite{Myers:1992}), and is argued mainly by simulations
to be a generic feature of long-range interacting systems in NESSs
\cite{Casetti:2014,Teles:2015}; here, we provide an analytic demonstration of the phenomenon.

To illustrate the generality of our method, we show in
Fig.~\ref{fig-ness-nemic} the results of adding a $\cos 2\theta$
interaction to the Kuramoto potential
($\widetilde{u}_1=0.3,\widetilde{u}_2=0.7,\widetilde{u}_{s>2}=0$). Also in this case we
see a perfect agreement of the theory with simulations. 

\begin{figure}
\centering
\includegraphics[width=75mm]{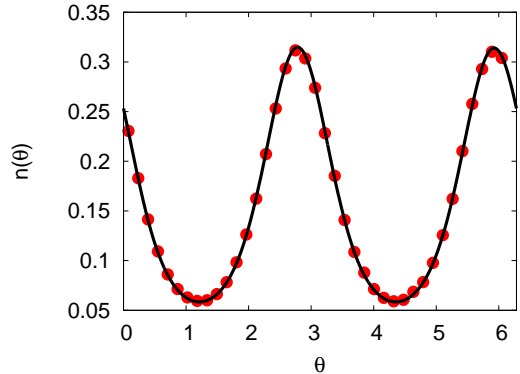}
\caption{Density $n(\theta)$ in the dynamics~(\ref{eom-scaled}) with
$\widetilde{u}_1=0.3$, $\widetilde{u}_2=0.7$, $\widetilde{u}_{s>2}=0$, a Gaussian ${\cal G}(\omega)$,
$m=0.25$, $T=0.25$, $\sigma=0.295$, $k_{\rm trunc}=4$.
Simulations (points) are for $N=10^6$; the theoretical predictions
are denoted by lines.}
\l{fig-ness-nemic}
\end{figure}

We now discuss the behavior of the truncation order $k_{\rm trunc}$
in computing the density $n(\th)$ as a function of $m$ at a
given representative $(\sigma,T)$; in particular, we point out that Eq. (\ref{bexpansion}) is an
asymptotic expansion in the inverse damping coefficient $\sqrt{m}$. Let us consider, e.g.,
parameter values used in this work, i.e., $\sigma = 0.295$, $T=0.25$ and $m=0.25$. For these values
$k_{\rm trunc}=18$ gives a perfect match with simulation results, as in the left panels of
Fig.~\ref{fig-ness-th} and Fig.~\ref{fig-ness-temp}. The match is no more perfect and gets worse and
worse on successively including more higher order terms in the expansion; in
this case, using the Borel method \cite{Hardy:1991} of summing a divergent
series circumvents the problem,
allowing to correctly compute $n(\theta)$ for a truncation order that in
principle could be arbitrarily large.
This is expected of an asymptotic expansion, and makes us conclude
that Eq. (\ref{bexpansion}) is an asymptotic expansion in $\sqrt{m}$. In
the Appendix, we give
some details on the Borel summation method.

\section{Conclusions}
We have proposed a novel method to compute the inhomogeneous NESS distribution of a wide class of mean-field
systems of rotators subject to quenched disordered external drive
and dissipation. We have demonstrated an excellent
agreement between simulations and theory for the noisy inertial Kuramoto
model of spontaneous collective synchronization, and for its two mode generalization. 

Our method is based on a series expansion of the stationary distribution function $f^{\rm syn}(\th,v,\omega)$.
First, the velocity dependence of the distribution is separated from the $(\th,\omega)$ dependence by an expansion in Hermite functions
of the velocity, with coefficients functions of $\th$ and $\omega$ (Eq. (\ref{f-expansion})); in turn, the latter
functions are expanded in powers of the inverse friction constant $\sqrt{m}$ (Eq. (\ref{bexpansion})). The second
expansion is asymptotic, but we have shown that, as is generally the
case with asymptotic series, we get the
``right'' sum by truncating at an appropriate order. Furthermore, as mentioned above and as detailed in the Appendix,
one can apply the Borel summation method to sum the expansion, a method
that often sums correctly an asymptotic series.
We stress that the appropriate order of truncation may be found even
without resorting to a comparison with simulation
data, since computation with a larger order of truncation leads to
numerical instabilities in the form of oscillations in the distribution, as
shown in the Appendix.

We note that our method does not determine if the computed inhomogeneous stationary solution is stable for given values of
the parameters. For this, it would be necessary to perform a stability
analysis, which for inhomogeneous solutions is much more complicated than
that for homogeneous solutions. However, by finding the inhomogeneous
solutions, one can theoretically determine the hystheresis loops associated with the presence of non-equilibrium first-order phase transitions
in the class of models we considered. In fact, the knowledge of the stability of the incoherent
$\theta$-independent solution (\ref{incsol}) as a function of the parameters \cite{Gupta:2014a}, and the determination of the
synchronized coherent solution, together allow to localize the hysteresis loops in the parameter space.

To sign off, we want to stress that the method can be applied also to
classes of models that generalize the one considered in this work \cite{Omelchenko:2012,Komarov:2014}.

\section*{Acknowledgments}
\addcontentsline{toc}{section}{Acknowledgments}
%%%%%%%%%%%%%%%%%%%%%%%%%%%%%%%%%%%%%%%%%%%%%%%%%%%%%%%%%%%%%%%%%%%%%
We thank M. Komarov and A. Pikovsky for discussions, and the Galileo Galilei Institute for Theoretical Physics,
Florence, Italy for the hospitality and the INFN for partial support during the completion of
this work. 
%%%%%%%%%%%%%%%%%%%%%%%%%%%%%%%%%%%%%%%%%%%%%%%%%%%%%%%%%%%%%%%%%%%%%

\appendix
\section*{Appendix: Convergence properties of the density expansion}
\l{appborel}
\setcounter{section}{1}
\addcontentsline{toc}{section}{Appendix: Convergence properties of the density expansion}
Consider an asymptotic power series in the real variable $x$,
\be\label{origseries}
A(x)=\sum_{k=0}^\infty a_kx^k,
\ee
and define the partial sum
\be
A_n(x) \equiv \sum_{k=0}^n a_k x^k.
\ee
Being asymptotic means that at any given $x\ne 0$, one has $A_n(x) \to \infty$
as $n \to \infty$. In this case, one might resort to the Borel summation method
by defining the Borel transform of $A(x)$ as \cite{Hardy:1991}
\be
{\cal B}A(t) \equiv \sum_{k=0}^\infty \fr{a_k}{k!}t^k.
\ee
If ${\cal B}A(t)$ converges for any positive $t$, or, if it converges for sufficiently
small $t$ to an analytic function that can be analytically continued to all $t>0$, and if
the integral
\be
\int_0^\infty \dd t~\exp(-t) {\cal B}A(tx)
\ee
exists and equals $A_B(x)$ (where the subscript $B$ stands for Borel),
then we say that the Borel sum of the series on the right hand
side of Eq.~(\ref{origseries}) is $A_B(x)$. It is not difficult to see that if the original series 
converges, i.e., if $\lim_{n \to \infty}A_n(x) = A(x) < \infty$, then $A_B(x) = A(x)$.
Applying the above formalism to Eq.~(\ref{bexpansion}), we get
\bea
b_{0B}(\th,\omega)&=&\int_0^\infty \dd t~\exp(-t) \sum_{k=0}^\infty
\fr{c_{0,k}(\th,\omega)}{k!}(t\sqrt{m})^k \nonumber \\
&=&\fr{1}{\sqrt{m}}\int_0^\infty \dd y~\exp(-y/\sqrt{m}) \sum_{k=0}^\infty
\fr{c_{0,k}(\th)}{k!}y^k.
\label{boreldensity}
\eea
The last integral is to be computed numerically. One has to truncate the
series at a certain order $k=k_{\rm trunc}$, and
to extend the integral over $y$ up to a given value $y_M$, which is chosen such that the integrand
becomes negligible for $y > y_M$. However, contrary to what happens in the
original series, we found that the sum in the last integral converges,
at least for all $y$-values smaller than $y_M$ that are necessary to compute the integral. We do not know the function to which our Borel
transform converges, and the corresponding radius of convergence, but the numerical results show that our
series is Borel summable. The left panel of Fig.~\ref{fig-borel-sum} shows the result of computing the density 
\be
n(\th)=\int_{-\infty}^\infty \dd \omega~ {\cal G}(\omega)b_{0B}(\th,\omega)
\ee
for the same conditions as in the left panel of Fig.~\ref{fig-ness-th}, by truncating the sum in
Eq. (\ref{boreldensity}) at $k_{\rm trunc}=38$; the plot coincides with the one shown in the left
panel of Fig.~\ref{fig-ness-th}.
\begin{figure}[ht]
\centering
\includegraphics[width=75mm]{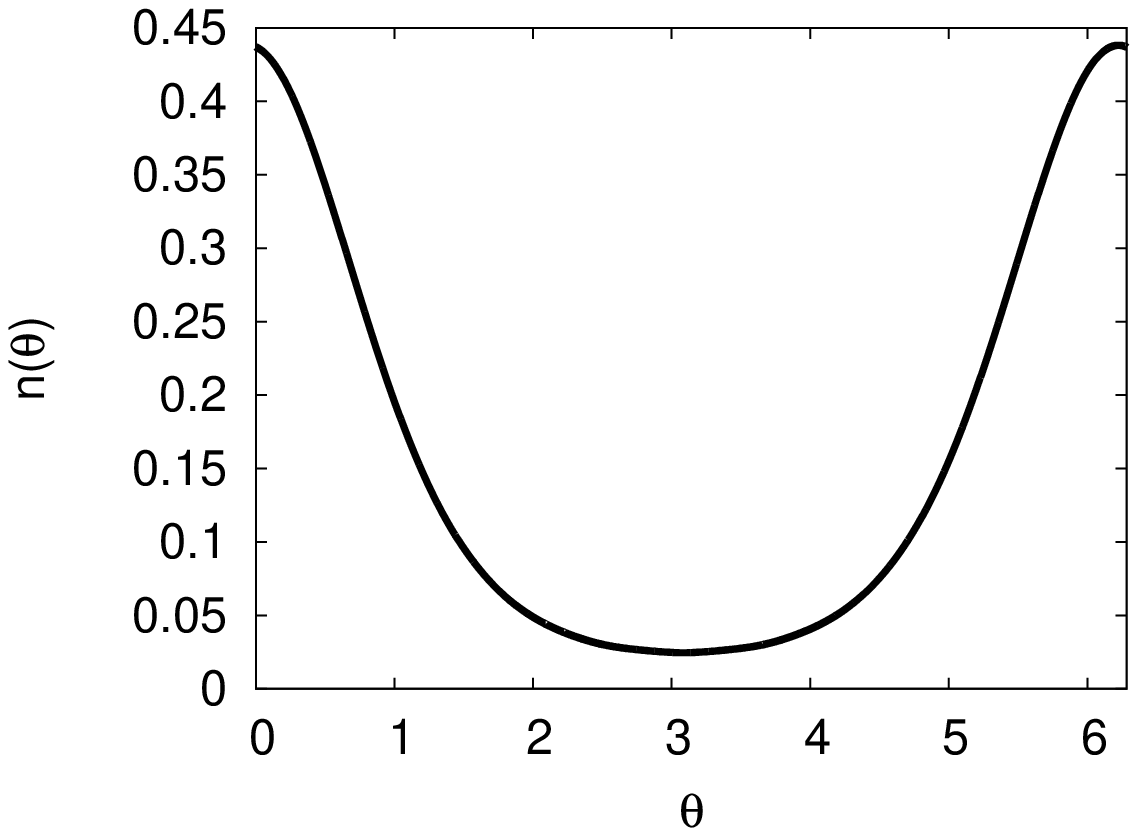}
\includegraphics[width=75mm]{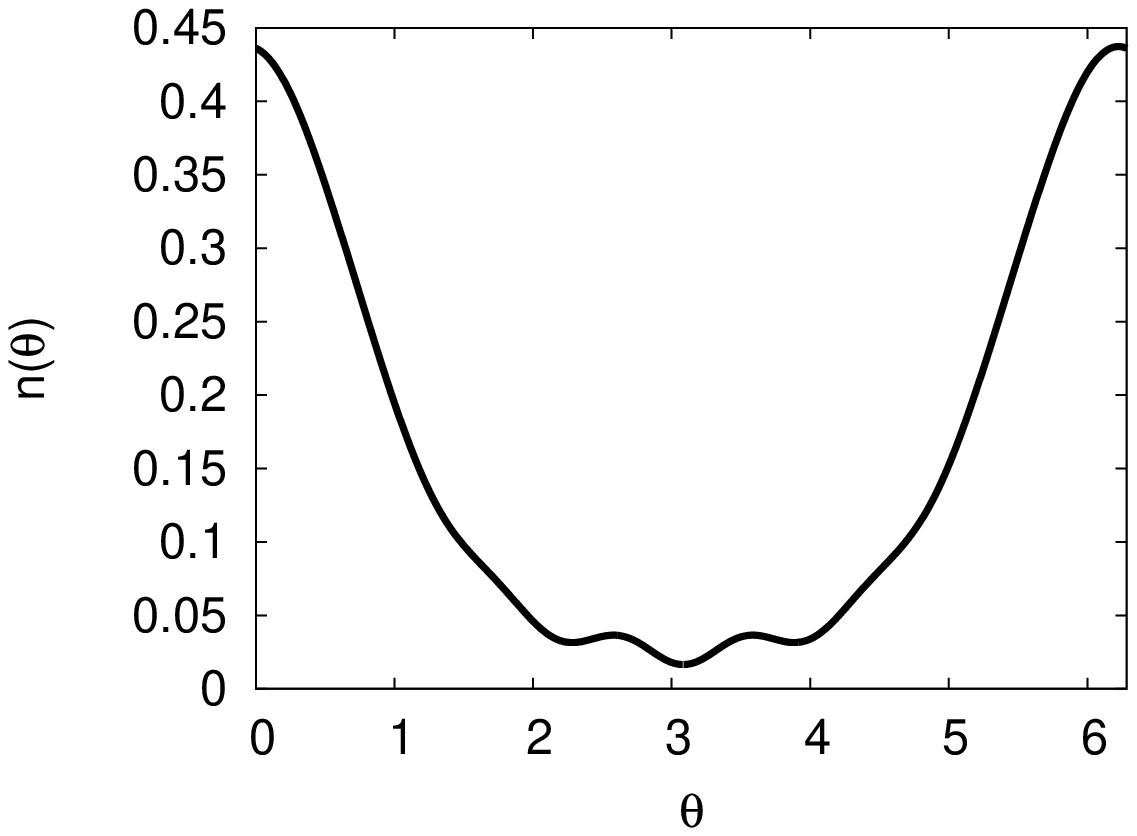}
\caption{Density $n(\theta)$ in the dynamics~(\ref{eom-scaled}), with
$\widetilde{u}_1=1$, $\widetilde{u}_{s>1}=0$, a Gaussian ${\cal G}(\omega)$, $m=0.25$, $T=0.25$, $\sigma=0.295$.
The left panel involves theoretical predictions using the Borel summation method with $k_{\rm trunc}=38$,
while the right panel involves those using direct summation with $k_{\rm trunc}=22$.}
\l{fig-borel-sum}
\end{figure}
On the other hand, summing the series~(\ref{bexpansion}) for $n=0$ without
resorting to the Borel summation method, and then computing the density $n(\th)$, the result in the right panel of
Fig.~\ref{fig-borel-sum} shows that already for truncation order $k_{\rm trunc}=22$
of the series, one observes instabilities that get worse and worse with
further increase of the truncation order (see Table \ref{table2} listing the
truncation order $k_{\rm max}$ as a function of $m$, for the same
representative $(\sigma,T)\equiv(0.295,0.25)$ as in Fig.~(\ref{fig-ness-th}),
up to which one observes a perfect agreement of the
density $n(\th)$ between theory and simulations). We conclude from this analysis that
the series~(\ref{bexpansion}), although asymptotic, is effectively summable by the
Borel summation method.

%\begin{table}[h]
%\centering
%\begin{tabular}{|c|c|}
%\multicolumn{2}{c}{} \\\hline
%$m$ & $k_{\rm max}$\\ \hline 
%0.0625 & 60 \\
%0.125 & 32 \\
%0.25 & 18 \\
%0.5 & 10 \\
%1.0 & 6 \\ \hline
%\end{tabular}
%\caption{For the dynamics, Eq.~(\ref{eom-scaled}), with
%$\widetilde{u}_1=1$, $\widetilde{u}_{s>1}=0$, and a Gaussian ${\cal G}(\omega)$,
%the table shows the maximum truncation order $k_{\rm max}$ in the computation of the density $n(\th)$ as a function of $m$
%at a given representative $(\sigma,T)\equiv(0.295,0.25)$ for which one observes a perfect agreement of the
%density $n(\th)$ in theory and simulations, as in Fig.~(\ref{fig-ness-th}).
%The agreement gets worse and worse on successively increasing truncation order beyond $k_{\rm max}$.}
%\l{table2}
%\end{table}

\begin{table}
\centering
\begin{tabular}{|c|c|c|c|c|c|}  \hline
$m$                    &$0.0625$   & $0.125$  & $0.25$   & $0.5$  & $1.0$ \\ \hline 
$k_{\rm max}$     &$60$ & $32$ & $18$ & $10$ & $6$  \\ \hline
\end{tabular}
\caption{For the dynamics, Eq.~(\ref{eom-scaled}), with
$\widetilde{u}_1=1$, $\widetilde{u}_{s>1}=0$, and a Gaussian ${\cal G}(\omega)$,
the table shows the maximum truncation order $k_{\rm max}$ in the computation of the density $n(\th)$ as a function of $m$
at a given representative $(\sigma,T)\equiv(0.295,0.25)$ for which one observes a perfect agreement of the
density $n(\th)$ in theory and simulations, as in Fig.~(\ref{fig-ness-th}).
The agreement gets worse and worse on successively increasing truncation order beyond $k_{\rm max}$.}
\l{table2}
\end{table}

%%%%%%%%%%%%%%%%%%%%%%%%%%%%%%%%%%%%%%%%%%%%%%%%%%%%%%%%%%%%%%%%%%%%%

\begin{thebibliography}{99}
\bibitem{note2} This ensures that the force on the $i$th rotator due to
the $j$th one is equal and opposite to that on the $j$th rotator due to the $i$th,
as required by the Newton's third law.
\bibitem{note3}
We will from now on take unity for the Boltzmann constant $k_B$.
\end{thebibliography}
\end{document}